\pdfoutput=1		
\documentclass[preprint,5p,times,twocolumn]{elsarticle}

\usepackage{siunitx,upgreek}
\usepackage{booktabs}
\usepackage{graphicx}
\usepackage{amsmath,amssymb}
\usepackage{color}
\usepackage{natbib}
\usepackage[colorlinks]{hyperref}
\usepackage{xspace}							
\graphicspath{{./figures/}}

\usepackage{framed} 		
\usepackage{multicol} 	
\usepackage{nomencl} 	
\makenomenclature

\renewcommand*\nompreamble{\begin{multicols}{2}}
\renewcommand*\nompostamble{\end{multicols}}

\newlength{\nomitemorigsep}
\usepackage{xpatch}
\usepackage{ifthen}
\renewcommand{\nomgroup}[1]{%
	\itemsep\nomitemorigsep%
	\ifthenelse{%
		\equal{#1}{G}%
	}{%
		\item[\textit{Greek Symbols}]%
		
	}{%
		\ifthenelse{\equal{#1}{S}}{%
			\item[\textit{Subscripts}]%
			
		}{}%
	}%
	\itemsep\nomitemsep
}

\renewcommand{\deg}{\,^\circ}

\providecommand{\e}[1]{\ensuremath{\times 10^{#1}}}
\newcommand{\pd}[2]{\frac{\partial #1}{\partial #2}}
\newcommand{\pdd}[2]{\frac{\partial^2 #1}{\partial #2^2}} 
\newcommand{\dpd}[2]{\dfrac{\partial #1}{\partial #2}}
\renewcommand{\d}[2]{\frac{\text{d} #1}{\text{d} #2}} 
\newcommand{\ol}[1]{\overline{#1}}
\renewcommand{\vec}[1]{\boldsymbol{#1}} 
\newcommand{\mat}[1]{\boldsymbol {\mathsf {#1}}} 

\newcommand{\Sh}{\mbox{\textit{Sh}}}  
\newcommand{\Sc}{\mbox{\textit{Sc}}}  
\newcommand{\Sr}{\mbox{\textit{Sr}}}  
\newcommand{\Pen}{\mbox{\textit{Pe}}}  
\newcommand{\Rey}{\mbox{\textit{Re}}}  

\newcommand{\ainv}{\alpha_{\text{inv}}}
\newcommand{\aexp}{\alpha_{\text{exp}}}
\newcommand{\aq}{\alpha_{\text{q}}}
\newcommand{\Sexp}{S_{\text{exp}}}
\newcommand{\sq}{s_{\text{q}}}
\newcommand{\Sq}{S_{\text{q}}}
\newcommand{\Sinv}{S_{\text{inv}}}
\newcommand{\ssob}{s_{\text{sob}}}
\newcommand{\Ssob}{S_{\text{sob}}}
\newcommand{\shsq}{\Sh^*_\text{q}}

\newcommand{\shexps}{\Sh^*_{\text{exp}}}
\newcommand{\klev}{k_\text{q}}

\newcommand{\betas}{\beta_S}
\newcommand{\betaa}{\beta_\alpha}
\newcommand{\alphao}{\alpha_0}

\newcommand{\lev}{L\'{e}v\^{e}que\xspace}
\newcommand{\sob}{Sobol\'ik\xspace}

\newcommand{\LADYF}{{\fontfamily{lmr}\selectfont\textup{\normalsize L\kern -.35em\lower -.5ex\hbox{\scriptsize A} \normalsize\kern -.4em\lower .2ex
			\hbox{\footnotesize D}\kern -.24em\lower .ex\hbox{Y}\kern -.1em
			\hbox{\scriptsize F}}}}

\begin{document}
\begin{frontmatter}
	\title{On the inverse problem for two-component unsteady wall shear rate measurements: Application to the electrodiffusion method}
	
	\author[label1]{Marc-\'Etienne Lamarche-Gagnon\corref{cor1}}
	\cortext[cor1]{Corresponding author}
	\ead{m-e.lamarche-gagnon@polymtl.ca}
	
	\author[label1]{J\'er\^ome V\'etel}
	
	\address[label1]{Department of Mechanical Engineering, \LADYF, Polytechnique Montreal, Montreal, Qc, H3T 1J4, Canada}

	\begin{abstract}
	The instantaneous two-dimensional wall shear rate is assessed through an inverse problem using mass transfer data from a three-segment electrodiffusion probe. The method is validated numerically in complex flow conditions involving (i) high amplitude periodic fluctuations on both wall shear rate magnitude and direction and (ii) direct numerical simulation (DNS) data from a turbulent three-dimensional channel flow. The approach is shown to outperform every other post-treatments available for mass transfer sensors, especially regarding its versatility and application range. The impact of the three-segment probe gap size is also examined numerically. 
	\end{abstract}
	
\end{frontmatter}

\begin{table*}[!t]   
\begin{framed}
\nomenclature[E]{$D$}{diffusion coefficient}
\nomenclature[E]{$A$}{area of the probe}
\nomenclature[E]{$d$}{diameter of a circular ED probe}
\nomenclature[E]{$f$}{frequency}
\nomenclature[E]{$F$}{Faraday constant}
\nomenclature[E]{$c$}{concentration}
\nomenclature[E]{$C$}{dimensionless concentration, see \eqref{eq:cdadim3D}}
\nomenclature[E]{$C_0$}{concentration in the bulk solution}
\nomenclature[E]{$k_q$}{\lev constant}
\nomenclature[E]{$k^*$}{constant in \eqref{eq:Shlev}, equals to $\Sh^*_\text{tot}$ when $\Pen\to\infty$}
\nomenclature[E]{$I$}{limiting current}
\nomenclature[E]{$J$}{reaction rate}
\nomenclature[E]{$\mat{J}$}{sensitivity matrix, see \eqref{eq:jac}}
\nomenclature[E]{$N$}{number of unknowns in the inverse problem}
\nomenclature[E]{$\vec{p}$}{vector of unknowns in the inverse problem}
\nomenclature[E]{$\Pen$}{P\'eclet number, see \eqref{eq:cdadim3D}}
\nomenclature[E]{$s$}{wall shear rate}
\nomenclature[E]{$S$}{dimensionless wall shear rate}
\nomenclature[E]{$\Sc$}{Schimdt number ($\nu/D$)}
\nomenclature[E]{$\Sh$}{Sherwood number}
\nomenclature[E]{$\Sh^*$}{modified Sherwood number $\left(\Sh\Pen^{-1/3}\right)$}
\nomenclature[E]{$\Sr$}{Strouhal number, see \eqref{eq:cdadim3D}}
\nomenclature[E]{$\vec{u}$}{velocity vector}
\nomenclature[E]{$t$}{time}
\nomenclature[E]{$x,y,z$}{streamwise, normal and spanwise coordinates}
\nomenclature[E]{$X,Y,Z$}{dimensionless coordinates, see \eqref{eq:cdadim3D}}

\nomenclature[G]{$\alpha$}{wall shear rate direction}
\nomenclature[G]{$\alpha_0$}{time average wall shear rate direction}
\nomenclature[G]{$\betaa$}{amplitude on $S$ for periodic flows, see \eqref{eq:Salpha}}
\nomenclature[G]{$\betas$}{amplitude on $\alpha$ for periodic flows, see \eqref{eq:Salpha}}
\nomenclature[G]{$\phi$}{phase shift between periodic $S$ and $\alpha$}
\nomenclature[G]{$\tau$}{dimensionless time, see \eqref{eq:cdadim3D} and \eqref{eq:tauInst}}
\nomenclature[G]{$\nu$}{kinematic viscosity}
\nomenclature[G]{$\zeta$}{attenuation ratio, see \eqref{eq:zeta}}

\nomenclature[S]{exp}{experimental or `true'}
\nomenclature[S]{sob}{\sob method}
\nomenclature[S]{q}{quasi-steady method}
\nomenclature[S]{$n$}{relative to unknown $n$}
\nomenclature[S]{$m$}{segment $m$ of a three-segment probe}

\printnomenclature
\end{framed}
\end{table*}

\section{Introduction}\label{sec:introduction}
Every wall confined flow is subject to wall shear stress, affecting the efficiency of many industrial systems such as pumps, turbines, heat exchangers or any application implying fluid circulation. Despite the considerable efforts in developing new methods over the years, measurement of wall shear stress remains a challenge, especially when both time and space resolution are required. Among the many available methods, the floating element is interesting due to the probe size and considerably large bandwidth \citep[sometimes up to 4\,kHz, see][]{padmanabhan1997}. However, spatial resolution of the floating element is generally limited by the overall electronic components surrounding the probe and those sensors can rarely measure at the same time the shear stress direction. The hot-film anemometer has also been widely used to assess wall shear rate in unsteady flows. On paper, hot-film frequency response stays flat up to a few kHz \citep{wietrzak1994wall}, but this value is largely weakened considering heat conduction through the wall, which also introduces a bias error. Regardless of the many adaptations developed over the last few decades (among other things: reduction of the substrate thermal conductivity, creation of a vacuum cavity below the sensor), this problem persists and accurate measurements under unsteady conditions are still an issue \citep{sheplak2004mems}. Natural convection caused by heat transfer to the fluid, known as the induced buoyancy, can also locally alter the flow conditions. \citet{he2011wall} attributed the large dispersions in their calibration data to this phenomenon, observing discrepancies as high as 10\,\% (turbulent pipe flow with $\Rey<18\,000$). Although flow direction can be assessed by arranging two or more hot-film sensors with different orientations, the poor probe sensitivity in the transverse direction limits its uses to small angles; such sensor configuration is usually reserved for detecting shear reversal rather than its direction \citep{bruun1995hot}.

The electrodiffusion (ED) method measures the electrolysis reaction rate between an electrode flush-mounted to a wall and a redox couple contained in the flow. The method is in many ways similar to the hot-film anemometry, where the local \textit{mass} transfer is measured instead of the heat transfer; the theory behind the two techniques overlaps in several aspects. Still, one major asset of the ED is the lack of heat loss to the wall, especially profitable in low convection flows. In their review of the wall shear stress produced by an impinging jet, \citet{phares2000wall} indeed concluded that the ED method provided `the most accurate data close to the stagnation point', as the ones from hot-film probes suffered from strong discrepancies. While the cutoff frequency associated with ED probes is rather low, adequate post-processing can correct the attenuation and phase shift of the sensor response in highly unsteady flows. In particular, by considering that the reaction at the probe interface is governed by the convection--diffusion equation, one can take advantage of the so-called \textit{inverse problem} to deal with the probe inertia. With this approach, the input wall shear rate is iteratively adjusted by solving the \textit{direct problem} (i.e. the convection--diffusion equation) until the numerical data converge to the experimental ones. According to \citet{rehimi2006inverse}, such method allows to accurately correct the probe response in high amplitude unsteady flows, including the case of shear reversal. The authors have demonstrated that this method outmatches every other post-treatments in two-dimensional flows. Considering its success, we propose an enhancement of the method adapted to three-dimensional flows able to capture the wall shear rate magnitude along with its direction in any unsteady flow when using a three-segment probe. The method is validated numerically for flows subjected to periodic and stochastic variations of the wall shear rate.

\begin{figure*}
	\centering
	\includegraphics[scale=0.95]{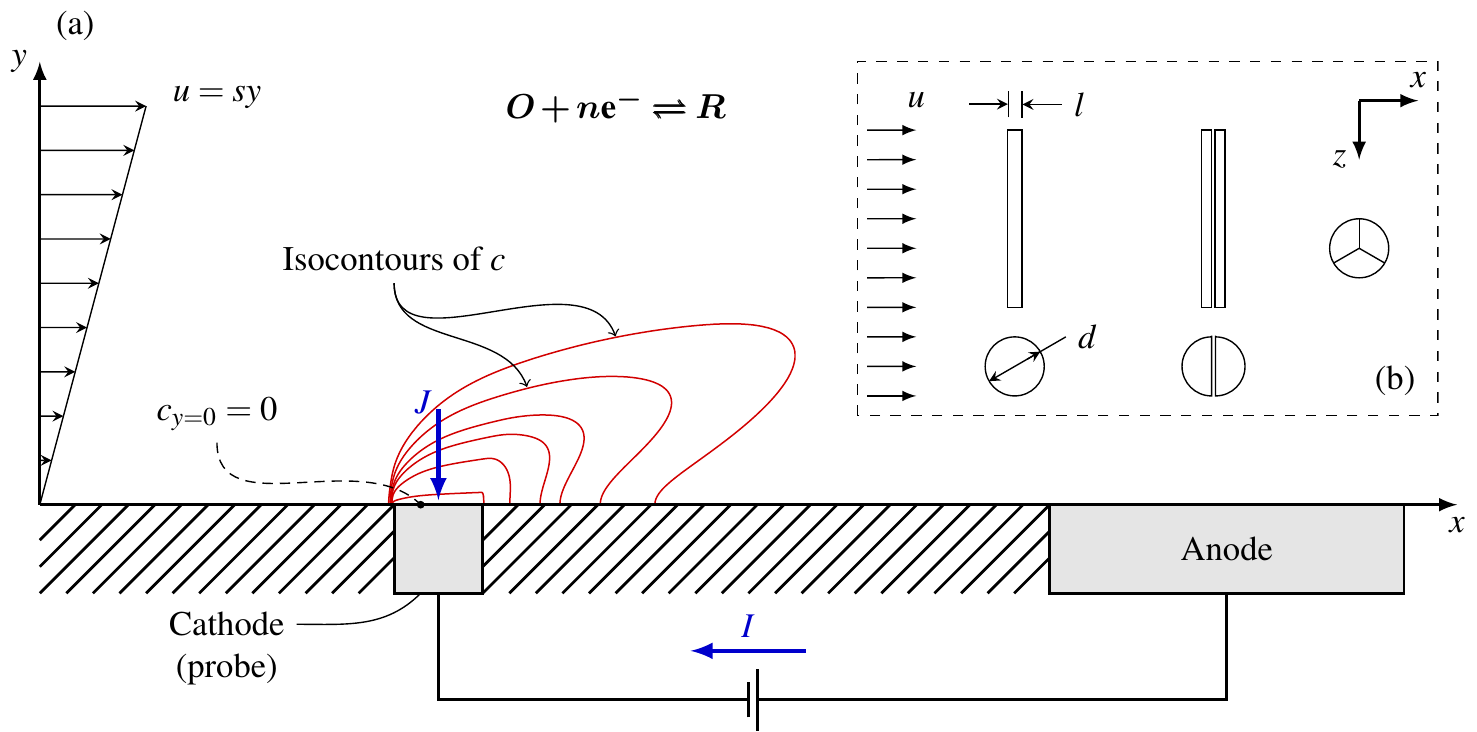}
	\caption{Principle of the electrodiffusion (ED) method. (a) Under constant voltage, a time varying current $I(t)$ flows throw the loop cathode-solution-anode, generating concentration gradients at the electrodes surface from $C_0$ in the bulk to a concentration $c=0$ at the probe--solution interface. (b) Typical probes, viewed from above. From left to right: single, double (sandwich) and three-segment probes.}
	\label{fig:1}
\end{figure*}

\subsection{Basic principles of the electrodiffusion method}
The ED method relies on the mass transfer between a redox couple ($O$--$R$) contained in a solution and two electrodes flush mounted to a wall (Fig.~\ref{fig:1}). The reaction is initiated by imposing a constant voltage between the anode and the cathode, where a concentration gradient gradually builds up at the electrodes--solution interface. The current $I$ flowing through the electrodes and electrolyte is then a function of the solution supply and is directly related to the wall shear rate $s$ under steady conditions. When streamwise convection is dominant, the relation $s\propto I^3$ typically holds over a certain voltage range, namely when the \textit{limiting current} condition is achieved \citep{hanr1996meas}; the electrode process then occurs at the maximal rate possible and the concentration at the probe surface is essentially null ($c_{y=0}=0$). Under this electrolysis process, mass transfer is manifested by an exchange of electrical charges between the $O$--$R$ species. While this transfer is normally assured by three methods (migration, diffusion and convection), a non-reactive or background electrolyte is usually added in excess to the solution so as to limit migration effects. The divergence of the resulting Nernst--Planck equation, which dictates the mass transfer at an electrode, results in the general convection--diffusion equation in absence of migration:
\begin{equation}
\pd{c}{t} + \vec{u}\cdot\nabla c = D\nabla^2 c,
\label{eq:CD}
\end{equation} 
with $D$ the diffusion coefficient. The relation between \eqref{eq:CD} and measures of $I(t)$ can also be derived from the Nernst--Planck equation. Under the assumption of a Nernst diffusion layer\footnote{At the electrode surface, a stagnant layer of thickness $\delta_c$ is assumed; in other words, convection is neglected in this area, resulting in a frozen diffusion layer \citep{bard2001electrochemical}.}, only the diffusion term remains and the flux or reaction rate $J$ at the probe can be written as
\begin{equation}
J_{y=0} = \frac{I}{nFA} = -\frac{1}{A}\iint_{A}D\left.\pd{c}{y}\right|_{y=0}dA,
\label{eq:fluxJ}
\end{equation}
with $n$, $F$, $A$ the number of electron(s) involved in the reaction, the Faraday number and the probe area, respectively. The reaction is then said to be diffusion-controlled.

The fundamental principle behind the ED method is that the reaction occurs very close to the wall owing to the thinness of the  concentration boundary layer (also called \textit{diffusion layer}) compared to the hydrodynamic one, so as to assume a linear velocity profile in the vicinity of the probe, $\vec{u}=\vec{s}y$, with $\vec{s}$ the two-dimensional wall shear rate. This is expected when the Schmidt number $\Sc=\nu/D$ is large\footnote{A typical value encountered is $\Sc\sim1000$.}, where $\nu$ is the kinematic viscosity. While two electrodes are needed for the process to occur, only the half-reaction arising at the working electrode will be of interest. The cathode is generally chosen as the working electrode, which will be referred to as the probe throughout the rest of the paper.

\subsection{Common post-treatment methods}

Various methods have been developed over the years to relate the measured current $I$ from an ED probe to the wall shear rate $s$ using assumptions on the flow or electrochemical conditions to model the process. When one only seeks a time averaged value, the \lev solution can be used, obtained from the analytical solution of the two-dimensional steady state convection--diffusion equation with linear velocity profile. The current can then be related to $s=\lVert\vec{s}\rVert$ through \eqref{eq:fluxJ}, where
\begin{equation}
I = \klev s^{1/3},
\label{eq:Ilev}
\end{equation}
with $\klev=0.80755nFAC_0l^{-1/3}D^{2/3}$ \citep{reiss1963expe} and $l$ is the rectangular probe size (cf. Fig.~\ref{fig:1}b). Equation \eqref{eq:Ilev} is also known as the \textit{quasi-steady method} and its validity is limited to slow time varying processes as the probe inertia is felt even at very low frequencies. The potential of ED probes to study wall turbulence first drove authors to develop models that would improve its frequency response, most of them based on the linearised turbulent fluctuations equation (among others: \cite{mitc1966stud,fort1971freq,mao1985use}). In particular, the ED probe transfer function derived by \citet{desl1990freq} can accurately correct this capacitive effect over a broad frequency range; its usage is however limited to small fluctuations on the wall shear rate as a result of the linear theory approximation. The most popular post-treatment for large amplitude unsteady flows is the one developed by \citet{sobolik1987simultaneous}, which was also derived a few years later in the work of \citet{wang1993approximate}. Commonly referred to as the \textit{\sob method}, it basically corrects the quasi-steady wall shear rate $\sq$ using its time derivative, namely
\begin{equation}
\ssob = \sq + \frac{2}{3}\chi\sq^{-2/3}\d{\sq}{t},
\end{equation}
with $\chi=0.80755^{-2}\pi^{-1}l^{2/3}D^{-1/3}$. \sob method is simple and very efficient to correct for the probe inertia, but it fails when dealing with flow reversal. Techniques have been developed to tackle this issue (in particular, the works of \citet{pedley1976heat} and \citet{menendez1985use}), but none of them can match the efficiency of the inverse method to evaluate the magnitude of the wall shear stress in high amplitude unsteady flows. When one also seeks the flow direction, either double or three-segment probes can be used (see Fig.~\ref{fig:1}b), while for the former sensitivity to transverse flow is poor and the use of the so-called `sandwich' probe (as first developed by \citet{son1969velo}) is thus normally limited to flow reversal detection in low-frequency oscillating flows. Turbulent fluctuations in the transverse direction were also extensively investigated by modeling the two-segment probe response  \citep[see][]{sirk1970limi, tour1978beha, desl2004near}, but the linear theory again involved limits such usage to low-amplitude perturbations and none have demonstrated the ability to measure the instantaneous two components of the wall shear rate. The three-segment sensor was first developed to fill this gap \citep{wein1987theory}. In steady or low-frequency flows, the probe can successfully detect local flow direction $\alpha$ with precision often better than $10\,^\circ$ with an appropriate calibration \citep{sobolik1990three}, but actually fails when unsteadiness is dominant due to the probe inertia (see Section~\ref{sec:valinv}). No direct method can currently correct the capacitive effect on each individual segment, thus explaining the three-segment sensor deficiency for instantaneous measurements. The inverse method was previously used with a sandwich probe by \citet{mao1992measurement}. The authors showed that shear reversal in a turbulent pipe flow is well captured by the method when using such a probe; however, no complementary measures were available to evaluate the actual precision of the method. Axial diffusion was also neglected in their model, hypothesis that is questionable near shear reversal phases and was blamed by the authors and others \cite{labraga2002wall} to cause discontinuities.

In this work, the inverse method is revisited. Taking advantage of the three-segment probe, both magnitude and direction of the wall shear stress can be assessed in high amplitude unsteady flows. This is possible by considering a second unknown in the inverse problem, which is the wall shear direction $\alpha$. In the following section, the direct problem will first be introduced, followed by a description of the inverse problem and algorithm. In Section~\ref{sec:numres}, simulated data are used to validate the inverse process in flows of growing complexity, including the case of a three-dimensional turbulent flow. Different probe geometries are also considered to investigate the impacts of the gap size and possible imperfections.

\section{Statement of the problems}
\subsection{Direct problem \label{sec:dirprob}}
\begin{figure*}
	\centering
	\includegraphics{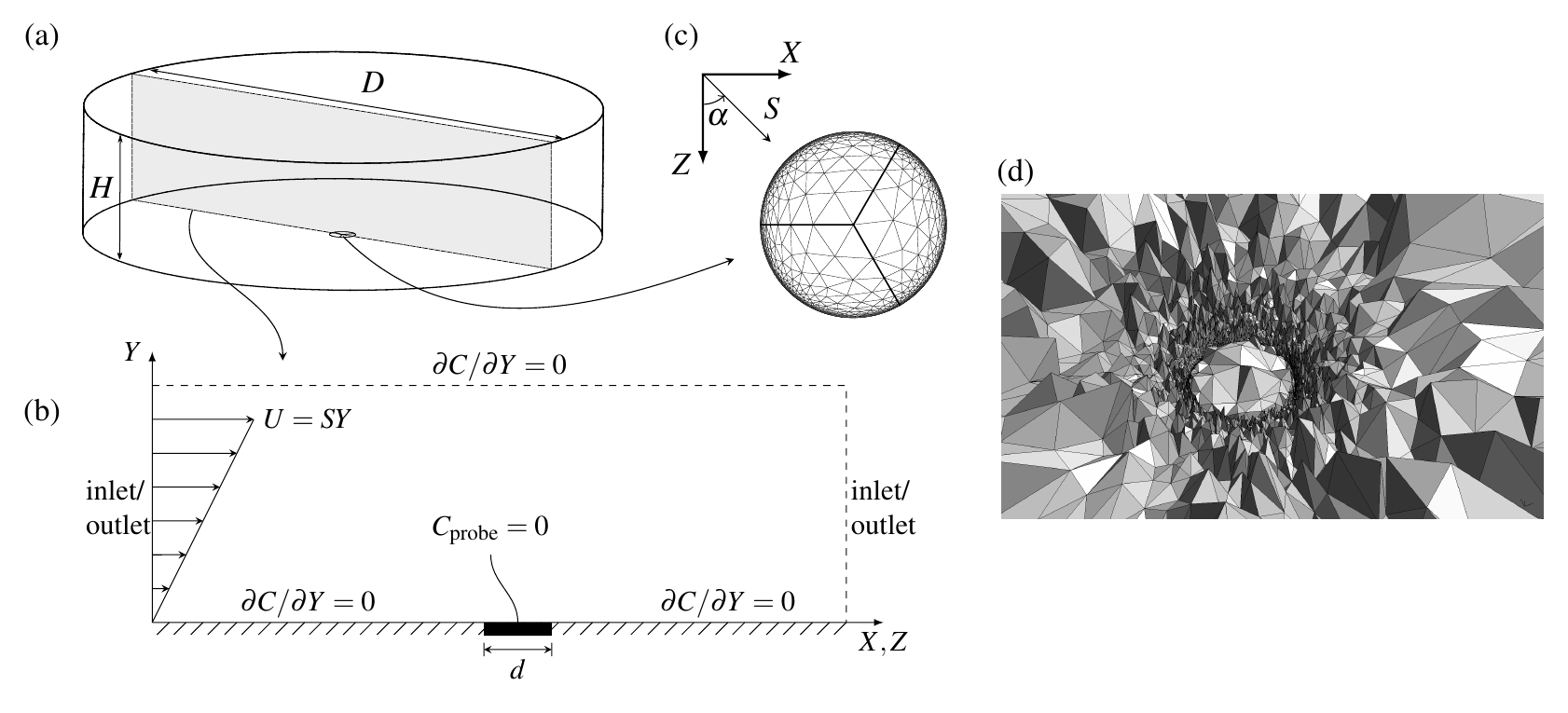}
	\caption{(a) Geometry and (b) boundary conditions used for the direct and the inverse problems. An overall diameter $D=50d$ and height $H=10d$ ensured that boundary conditions would not interfere with the probe reaction, even in cases of flow reversal. The inlet/outlet switches between Dirichlet ($C=1$) and Neumann ($\partial C/\partial n=0$) conditions depending on the flow direction $\alpha$. (c) Discretized null-gap three-segment probe (mesh G0 in Section \ref{sec:probdis}). Usage of small triangles near the probe edges are twofold: they help to manage the local strong concentration gradients, but they also serve for numerical stabilization of the problem without the need for a stabilization scheme. (d) Cut view in the $x-z$ plane of a typical mesh used. The three-segment probe is surrounded by the smallest elements.}
	\label{fig:2}
\end{figure*}
The three-dimensional convection--diffusion equation with a parallel linear velocity profile $\vec{u}=\vec{s}y$ is considered. In the presence of periodic flow fluctuations characterized by a frequency $f$, the dimensionless form of \eqref{eq:CD} can be written as
\begin{multline}
\Sr\pd{C}{\tau} +SY\left(\sin\alpha\pd{C}{X} + \cos\alpha\pd{C}{Z}\right) = \\ \Pen^{-2/3}\left( \pdd{C}{X} + \pdd{C}{Z}\right)  + \pdd{C}{Y},
\label{eq:cdadim3D}
\end{multline}
using the following dimensionless variables
\begin{gather}
\begin{array}{c}
X = \dfrac{x}{d},  \qquad  Z = \dfrac{z}{d},  \qquad  Y = \dfrac{y}{\delta},  \qquad  \tau = tf,  \\ \vspace{-5pt}\\
\Pen = \dfrac{\ol{s}d^2}{D},  \qquad  \Sr = \dfrac{fd^{2/3}}{\ol{s}^{2/3}D^{1/3}},  \qquad  S = \dfrac{s}{\ol{s}},  \qquad  C = \dfrac{c}{C_0},
\end{array}
\label{eq:adim}
\end{gather}
where $\Sr$, $\Pen$ will be respectively referred to the Strouhal and P\'eclet numbers, while $\delta=d^{-1}\Pen^{1/3}$ is the diffusion layer thickness under Nernstian approximation and $\ol{(\sim)}$ refers to a time averaged quantity over one period. The direct problem consists of solving \eqref{eq:cdadim3D} for given $S$ and $\alpha$ using the domain and boundary conditions synthesized in Figs.~\ref{fig:2}(a,b). A mix Dirichlet--Neumann condition (referred to as `inlet/outlet') is applied on the boundary $X^2+Z^2 = D^2/4$, which imposes $\partial C/\partial n=0$ when $\vec{u}\cdot\vec{n}>0$ (outflow) and $C=1$ otherwise. The dimensionless flux at the probe surface can be represented by the Sherwood number, defined as
\begin{equation}
\Sh = \frac{Jd}{C_0D}
\label{eq:Shdef}
\end{equation}
which becomes, using \eqref{eq:fluxJ} and \eqref{eq:adim} for the segment $m\in\{0,1,2\}$ of a three-segment probe,
\begin{equation}
\Sh_m=\frac{\Pen^{1/3}}{A}\iint_{A_m}\left.\pd{C}{Y}\right|_{Y=0}dA
\label{eq:Shseg}
\end{equation}
with $A_m$ the area of the discretized segment $m$. Often, one is more interested in the more convenient modified Sherwood number, simply defined as $\Sh^*=\Sh\Pen^{-1/3}$. The problem is solved using the open-source finite element equation solver \texttt{FreeFem++} \citep{MR3043640} with the use of P1 and P2 tetrahedral elements. Considering the discontinuity of the boundary conditions at the probe edges, anisotropic triangles were opted to discretize the probe surface (see Fig.~\ref{fig:2}c) which greatly reduced the computational cost and helped to stabilize the problem. In fact, solution of the convection--diffusion equation is subject to spurious oscillations at high $\Pen$. Using elements of small size thus eliminates the need for classical upwinding techniques such as SUPG methods; those inevitably add diffusion contributions in the vicinity of the probe, which here is undesirable considering that diffusion is the actual reaction to measure. The form of \eqref{eq:cdadim3D}, being stretched in the $y$ direction, also allowed the use of anisotropic tetrahedral elements. Such effect is observed in Fig.~\ref{fig:2}(d). Metrics associated with the concentration field surrounding the probe were used to evaluate the anisotropy parameters and construct the meshes, thanks again to \texttt{FreeFem++}. The time derivative was discretized using a backward differentiation formula (BDF) scheme of first and second orders.

\subsection{Inverse problem \label{sec:invpro}}
The time evolution of the \textit{true} (indicated with the subscript `exp') wall shear rate magnitude $\Sexp(\tau)$ and direction $\aexp(\tau)$ are to be found, with $\shexps(\tau)$ the only known information from, say, three-segment probe mass transfer measurements. For convenience, let $M_{m,i}=\Sh^*_{\text{exp},m,i}$ be the $\Sh^*$ measured by segment $m$ at time step $i$. The inverse method iteratively solves \eqref{eq:cdadim3D} by correcting the input values $S_i$ and $\alpha_i$ from time step $i$ to $T$ until a certain tolerance \texttt{tol} is reached. Then, one can expect the numerical conditions to represent the experimental ones well and so $S_i=S_{\text{exp},i}$, $\alpha_i=\alpha_{\text{exp},i}$. One of the most important aspects in a multidimensional inverse problem is the choice of the correction to be applied, which will affect both the convergence speed and stability of the algorithm, but also the smoothness and accuracy of the solution. For the case being, three $M_m$ measures are known at each time step and two variables are to be found. Hence, instead of using a classic Newton method, a Gauss--Newton algorithm is proposed. Let's consider the residual $r_m=M_m-\Sh^*_m$  at time step $i$ and the vector of unknowns $\vec{p}=[S,\alpha]$. Gauss--Newton algorithm will minimize the function $f=\sum r_m^2$ by using the following correction on vector $\vec{p}$:
\begin{equation}
\vec{p}^{(j+1)} = \vec{p}^{(j)} - (\mat{J}'\mat{J})^{-1}\mat{J}'\vec{r}^{(j)}
\label{eq:GN}
\end{equation}
with
\begin{equation}
\mat{J}=
\begin{bmatrix} \vspace{.3em}
\dpd{r_0}{S} & \dpd{r_0}{\alpha} \\ \vspace{.3em}
\dpd{r_1}{S} & \dpd{r_1}{\alpha} \\
\dpd{r_2}{S} & \dpd{r_2}{\alpha}
\end{bmatrix}
=
-
\begin{bmatrix} \vspace{.3em}
\dpd{\Sh^*_0}{S} & \dpd{\Sh^*_0}{\alpha} \\ \vspace{.3em}
\dpd{\Sh^*_1}{S} & \dpd{\Sh^*_1}{\alpha} \\
\dpd{\Sh^*_2}{S} & \dpd{\Sh^*_2}{\alpha}
\end{bmatrix}
\label{eq:jac}
\end{equation}
the jacobian matrix for iteration $j$. $\mat{J}$ can be interpreted as the sensitivity matrix, that is the sensitivity of the segment Sherwood number $\Sh^*_m$ to a variation of the parameter $p_n$, $n\in\{0,1\}$. One usually wants the magnitude of coefficients $J_{mn}$ to be as large as possible to increase accuracy in the estimates and reduce the effect of measurement errors. Hence, some smoothing methods exist to modify the matrix $\mat{J}$ when dealing with noisy experimental data so as to maximize the product $\mat{J'}\mat{J}$ \citep{ozisik2000inverse} or sometimes to remove singular values that make $\mat{J}$ ill-conditioned, using for instance SVD or other filtering methods \citep{shen2002solu,beck2016inve}. Nevertheless, such treatments were not needed and are not employed in the present paper. 

The traditional method to evaluate the sensitivity coefficients $J_{mn}$ in the one-dimensional inverse problem (i.e. when only the wall shear rate magnitude $S$ is sought) is with the finite difference approximation \citep[see for example][]{rehimi2006inverse} 
\begin{equation}
\pd{\Sh^*}{S} = \frac{\Sh^*(S+\epsilon S) - \Sh^*(S-\epsilon S)}{2\epsilon S},
\end{equation}
with $\Sh^*(S+\epsilon S)$ the Sherwood number under a wall shear rate $(1+\epsilon)S$, with $\epsilon\sim1\e{-5}$. Considering the present two-dimensional problem with $N=2$ unknowns, this process would require to solve \eqref{eq:cdadim3D} $2N$ additional times. A procedure also encountered in inverse problems is the use of sensitivity equations derived from the direct problem \citep{ozisik2000inverse}, obtained here by differentiating \eqref{eq:cdadim3D} with respect to $S$ and $\alpha$, namely
\begin{subequations}
	\begin{multline}
	\Sr\pd{C_1}{\tau} + SY\left(\sin\alpha\pd{C_1}{X} + \cos\alpha\pd{C_1}{Z}\right) = \Pen^{-2/3}\left( \pd{C_1}{X} + \pd{C_1}{Z}\right) \\  
	+ \pd{C_1}{Y}	- Y\left(\sin\alpha\pd{C}{X} + \cos\alpha\pd{C}{z}\right)
	\label{eq:C1}
	\end{multline}
	and
	\begin{multline}
	\Sr\pd{C_2}{\tau} + SY\left(\sin\alpha\pd{C_2}{X} + \cos\alpha\pd{C_2}{Z}\right) = \Pen^{-2/3}\left( \pd{C_2}{X} + \pd{C_2}{Z}\right) \\  
	 + \pd{C_2}{Y} - SY\left(\cos\alpha\pd{C}{X} - \sin\alpha\pd{C}{z}\right),
	\label{eq:C2}
	\end{multline}
	\label{eq:C1C2}
\end{subequations}
with $C_1=\partial C/\partial S$ and $C_2=\partial C/\partial \alpha$. The sensitivity coefficients of \eqref{eq:jac} are accordingly derived by differentiating \eqref{eq:Shseg} and thus
\begin{equation}
J_{mn} = \frac{1}{A}\iint_{A_m}\left.\pd{C_{n+1}}{Y}\right|_{Y=0}dA.
\end{equation}
Such a procedure is in many ways more appealing than the finite difference one considering that the computation of $N$ additional equations is required instead of $2N$ and no parameter $\epsilon$ needs to be defined. The major benefit still arises from the fact that \eqref{eq:C1} and \eqref{eq:C2} are actually the same equation as \eqref{eq:cdadim3D} added with a known source term, given that $C$ was already solved for. Hence, no matrix reconstruction is needed by the finite element solver when solving \eqref{eq:C1C2}, reducing computation time of each sensitivity equation by a factor $\sim10$.

\begin{figure*}
	\centering
	\includegraphics{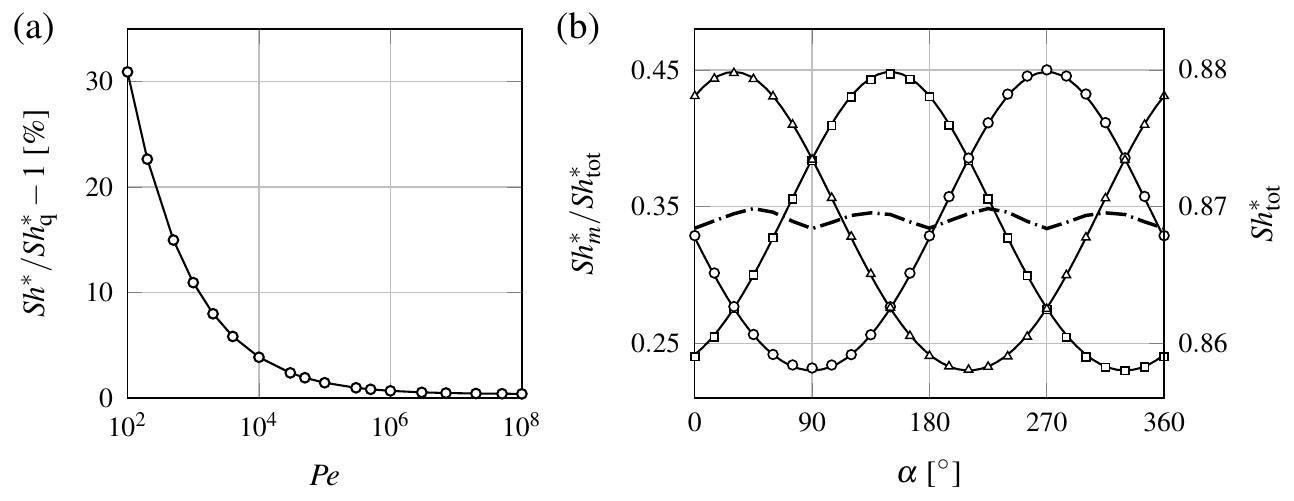}
	\caption[Validation of the direct problem]{Validation of the direct problem. In (a), solution of the stationary case for increasing values of $\Pen$ with $S=1$ indicates that discrepancies with \lev solution \eqref{eq:Shlev} is below 1.3\,\% (0.5\,\% using P2 tetrahedral elements) for $\Pen>5\e{6}$. (b) The directional characteristics (marks, where symbols refer to distinct segments) of the three-segment probe shown in Fig.~\ref{fig:2}(c) are compared to the modeled ones 
	(solid lines, see \cite{wein1987theory}). Also shown is the variation of the total Sherwood number $\Sh^*_\text{tot}=\sum\Sh^*_m$ over a complete rotation of the probe (dashed--dotted line), which is below $0.1\%$ of the mean value.}
	\label{fig:3}
\end{figure*}

The following iterative procedure is then proposed, given the starting guesses $\vec{p}^{(0)}$ from, for instance, the quasi-steady solutions. For each time step $i<T$,
\begin{enumerate}[(i)]
	\item solve the direct problem (cf. Section~\ref{sec:dirprob}) for time $\tau_i$ using $S=p_0^{j}$ and $\alpha=p_1^{j}$ and compute $f$;
	\item if $f<\texttt{tol}$, return to (i) for time $\tau_{i+1}$; otherwise
	\item solve the sensitivity equations \eqref{eq:C1C2} and compute $\mat{J}$.
	\item evaluate the new guesses $\vec{p}^{(j+1)}$ with \eqref{eq:GN} and return to (i) for iteration $j+1$.
\end{enumerate}
Note that for stabilization purposes, it is preferred to solve the first time step in steady condition ($\Sr=0$). Convergence speed is also very sensitive to the initial guesses, especially for $\alpha$, and the quasi-steady estimates are often not the best choices at high $\Sr$ number. For the one-dimensional inverse problem, \citet{rehimi2006inverse} suggest supposing that $S(\tau)=a\tau^2+b\tau+c$ on a short time interval and estimate the coefficients using the Levenberg--Marquardt method to ensure the start-up stability. For the present case, a line search method is coupled to the Gauss--Newton algorithm to ensure boundness of the process. When such approach is not sufficient (for instance, in some high-frequency cases), the safer but costlier conjugated gradient method is used instead of \eqref{eq:GN} to evaluate the corrections on $\vec{p}$. Most numerical methods and algorithms were implemented using procedures found in \cite{press2007numerical}. Moreover, a zero gradient verification is often preferred for the convergence criterion of step (ii), especially for high-frequency cases where the sensitivity coefficients are damped. As the main steps of the inverse process are very similar to the one-dimensional case, the reader is referred to other exhaustive sources for more details \citep[see][]{mao1991analysis,maquin99,ozisik2000inverse,rehimi2006inverse}. However, note that compared to most previous methods, a few iterations (denoted with superscript $(j)$) per time step are performed. Overall, this ensures better convergence properties.

\section{Results and discussion\label{sec:numres}}
Validation of the direct problem is first performed by analyzing the stationary case, i.e. equation \eqref{eq:cdadim3D} with $\Sr=0$. Considering a perfectly circular three-segment probe with negligible gaps (see Fig.~\ref{fig:2}c) and equivalent length $l_\text{eq}=0.81356d$ \citep{hanr1996meas}, the non-dimensional variant of the \lev solution \eqref{eq:Ilev} is formulated as
\begin{equation}
\shsq=k^*S^{1/3},
\label{eq:Shlev}
\end{equation} 
with $k^*=0.86505$, which value is expected when the tangential and transverse diffusion terms in \eqref{eq:cdadim3D} are negligible, that is for high P\'eclet numbers. As observed in Fig.~\ref{fig:3}, discrepancies with \eqref{eq:Shlev} is negligible when $\Pen\gtrsim1\e6$. To assess the direction $\alpha$ using a three-segment probe, its directional characteristics are required, obtained when performing the so-called directional calibration (as modeled by \citet{wein1987theory}). Under a steady flow, the probe signal is recorded while being gradually rotated; then, each segment relative signal $\Sh^*_m/\Sh^*_\text{tot}$ can be represented with a Fourier series expansion in $\alpha$ to deal with the probe geometry and imperfections. Such calibration was performed numerically to validate the directional characteristics of the discretized three-segment probe. Only small deviations are observed in Fig.~\ref{fig:3}(b) with the modeled characteristics of the perfect sensor \citep{wein1987theory}.

\subsection{Validation of the inverse problem \label{sec:valinv}}
\begin{figure*}[t]
	\centering
	\includegraphics[scale=0.95]{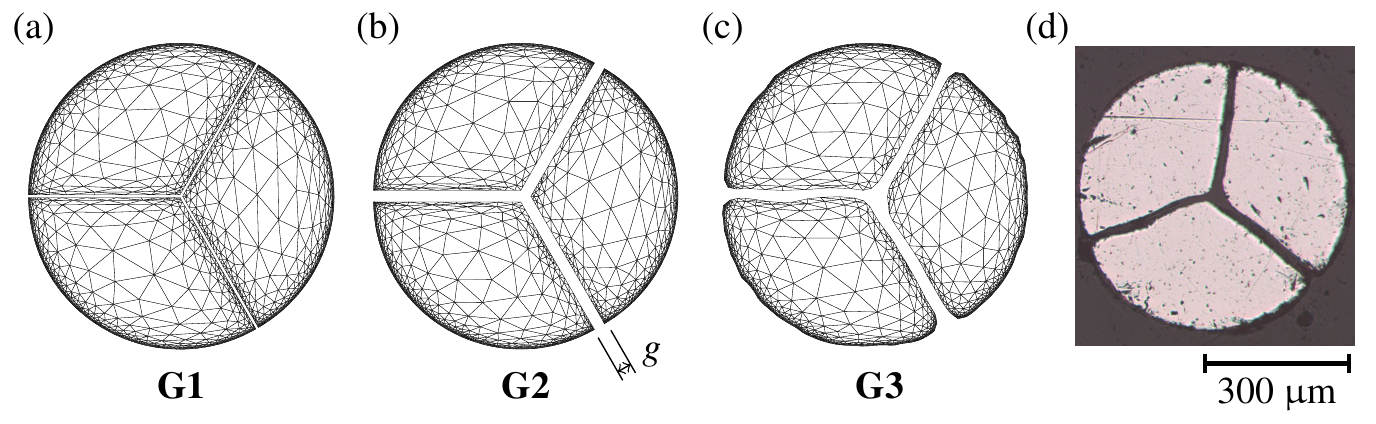}
	\caption[Gap effect]{Different geometries used to discretize the three-segment probe. (a) Smallest gap with a size equivalent to $g/4$; (b) largest gap, with the size $g$ close to the real probe G3; (c) discretization of the real probe geometry, which was shaped using a contour detection algorithm on (d), the optical microscope photograph of a real three-segment probe. All meshes were constructed so the total area of the three segments would respect $A=\pi/4$, that is the area of the equivalent null-gap probe G0 (Fig.~\ref{fig:2}b) with $d=1$.}
	\label{fig:9}
\end{figure*}

The use of the null-gap three-segment probe only being theoretical, a sensor with small gaps (see Fig.~\ref{fig:9}a) is used in the following section. Nevertheless, the use of the null-gap geometry in the previous section was necessary for validation purposes, considering that the \lev solution is not valid with an imperfect disk. Note that the mesh generation procedure was the same for all geometries. 

Validation of the inverse algorithm was accomplished by first simulating data using the direct problem for various flow parameters and then applying the inverse procedure of Section~\ref{sec:invpro}. Test cases are here based on the periodic fluctuation of both $S$ and $\alpha$, as per the following equations:
\begin{subequations}
	\begin{align}
	S(\tau) &= 1 + \betas\sin(2\pi\tau), \label{eq:SalphaA}\\
	\alpha(\tau) &= \alphao + \betaa\sin(2\pi\tau + \phi). 
	\end{align}
	\label{eq:Salpha}
\end{subequations}

\begin{table}[h!]
	\centering
	\caption{Parameters used in the simulations, as per equations \eqref{eq:Salpha}. }
	\begin{tabular*}{0.4\textwidth}{@{\extracolsep{\fill}} cccccc}
		\toprule
		Case	& $\Sr$ & $\betas$ & $\betaa$ & $\alphao$ & $\phi$	\\
		\midrule
		0	&	1.5	&	0.5	&	0	&	$\pi/3$	&	0	\\
		1	&	0.1	&	1.5	&	0	&	$\pi/3$	&	0	\\
		2	&	1.5	&	1.5	&	0	&	$\pi/3$	&	0	\\
		3	&	1.5	&	0.5	&	$\pi/4$	&	$\pi/2$	&	$\pi/6$	\\
		4	&	2		&	0.9	&	$2\pi/3$	&	$\pi/2$	&	$\pi/6$	\\
		5	&	0.5	&	0.5	&	$\pi/4$	&	$\pi/2$	&	$\pi/6$	\\
		\bottomrule
	\end{tabular*}
	\label{tab:parnum}
\end{table}
\begin{figure}[h!]
	\centering
	\includegraphics{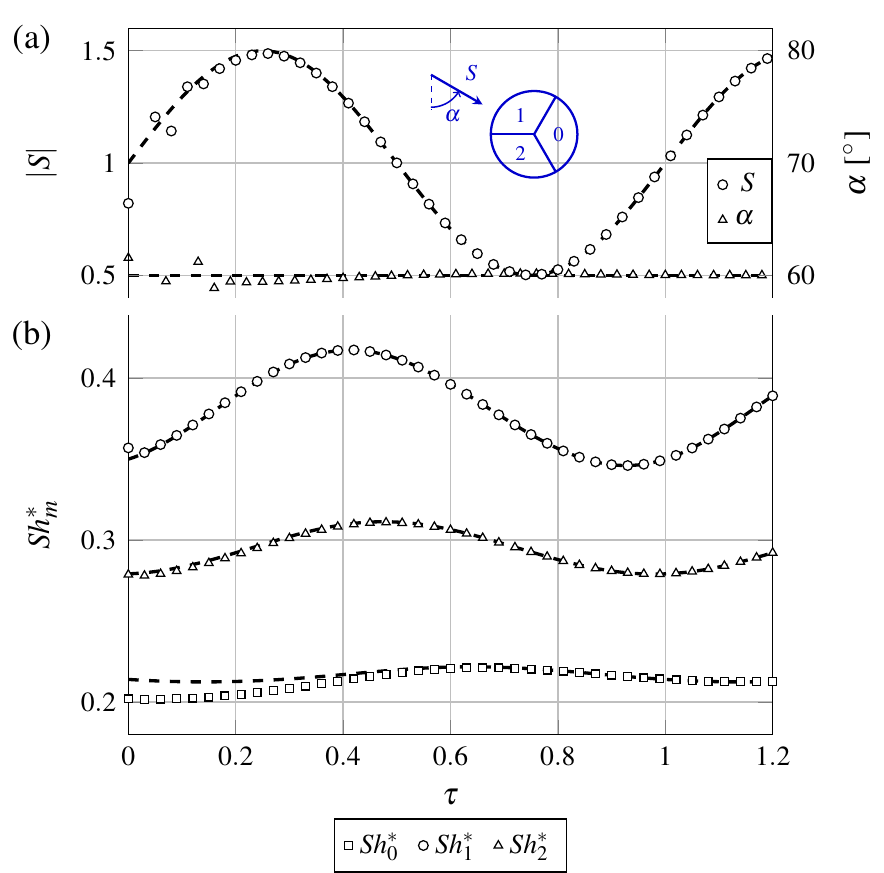}
	\caption[Inverse results 1]{Results of the inverse algorithm for constant direction $\alpha$, moderate amplitude and high-frequency fluctuation on $S$ (case 0, see Table~\ref{tab:parnum}). (a) After less than one period, $S$ and $\alpha$ converge on the true values (indicated by the dashed lines). Slower convergence for $\alpha$ could be explained by the lesser sensitivity of the back segment; convergence of both $\alpha$ and $\Sh^*_0$ appear in fact to be linked as observed in (b).}
	\label{fig:4}
\end{figure}

Simulations parameters are detailed in Table~\ref{tab:parnum}. For all cases, a P\'eclet number $\Pen~\to~\infty$ was used to offer the best comparison with other post-treatment methods since all are based on the \lev solution. P1 (linear) tetrahedral elements instead of P2 were employed for the tests. This does not affect the results as the same mesh and element type are used in both direct and inverse problems. Main drawback concerns the sensitivity equations, resulting in a less sensitive process and may thus imply one or two additional iterations for convergence; still, this greatly reduces computational costs ($\sim 10$ times faster). Note that the same frequency was used for $S(\tau)$ and $\alpha(\tau)$ in \eqref{eq:Salpha} to simplify the first step validation. This, however, does not limit the proposed algorithm as it will be exposed in Section \ref{sec:nonPer}.

\begin{figure*}[h!]
	\centering
	\includegraphics{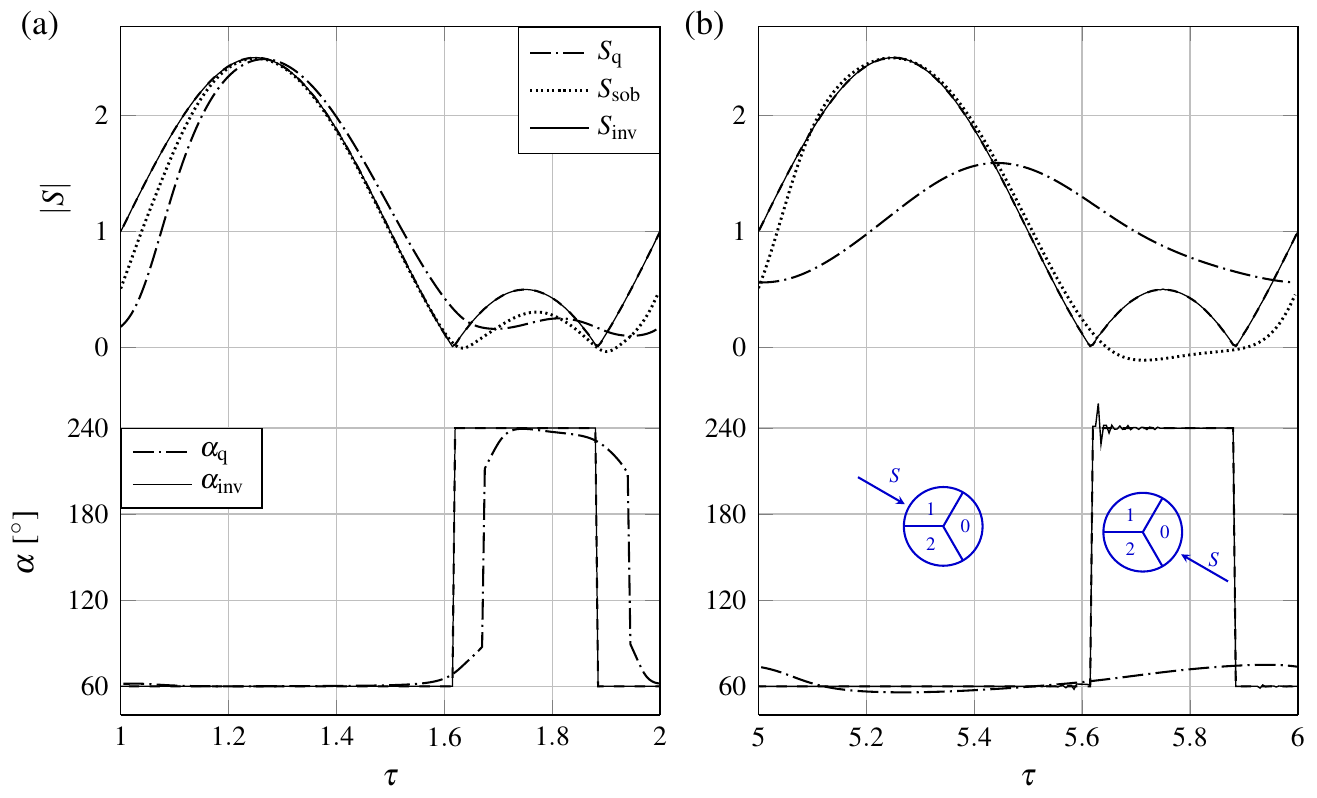}
	\caption[Inverse results 2]{Comparison of different post-treatment methods in two flows involving shear reversal. (a) In the low-frequency case 1 (see Table~\ref{tab:parnum}), the combined information of $\Ssob$ and $\aq$ still gives a decent approximation of the true wall shear rate, although the reversal period is out of phase; (b) at larger $\Sr$ (case 2), no valuable information can be retrieved from $\Ssob$ in the reversal period, while the quasi-steady method $\aq$ does not detect at all the shear reversal. In both cases, the inverse method shows almost perfect results. Only the last period of the inverse process is shown; two or more periods are sometimes needed for a suitable convergence depending on cases and initial guesses. Note that only the magnitude of $S$ is shown for visualization; when $S$ becomes negative (in cases with $\betas>1$), a $180\deg$ offset is added to $\alpha$.}
	\label{fig:5}
\end{figure*}

The inverse method efficiency and convergence speed can first be appreciated in Fig.~\ref{fig:4} for the constant direction, moderate shear fluctuation amplitude and high-frequency case 0, where only a few time steps are needed to converge on both the imposed values of $S$ and $\alpha$. Note that only part of the time steps used in the simulations are shown in all figures. In Fig.~\ref{fig:4}(b), one can notice that more time steps are needed for $\Sh^*_0$ to properly converge; being in the wake of the other two segments, the sensitivity coefficients related to this segment are lower. While this affects both sensitivity on $S$ and $\alpha$, the latter is more apparent considering the constant direction. 

Since all post-treatment methods give good results for low $\Sr$ without shear reversal, the following examples are limited to flows exhibiting either high $\Sr$, shear reversal, periodic fluctuation of $\alpha$ ($\betaa\neq0$) or a combination of the former, namely situations where both quasi-steady and \sob solutions fail. Also note that for the remainder of the paper, only results from the last period of periodic processes are shown, after which no further convergence improvements were observed. Cases 1 and 2 (Fig.~\ref{fig:5}) involve shear reversal in low- and high-frequency ranges, respectively. Although \sob method theory is limited for flows with constant direction, the combine information of $\Ssob$ and $\aq$ actually suggests the presence of shear reversal, yet suffering from an appreciable phase lag as observed in Fig.~\ref{fig:5}(a). At higher $\Sr$ however, the quasi-steady method cannot detect at all the reversal period and no valuable information is neither obtained from $\Ssob$ under such conditions (Fig.~\ref{fig:5}b). One can notice that $\Sinv$ and $\ainv$ converge almost perfectly on the imposed fluctuations in both cases. 

\begin{figure}
	\centering
	\includegraphics{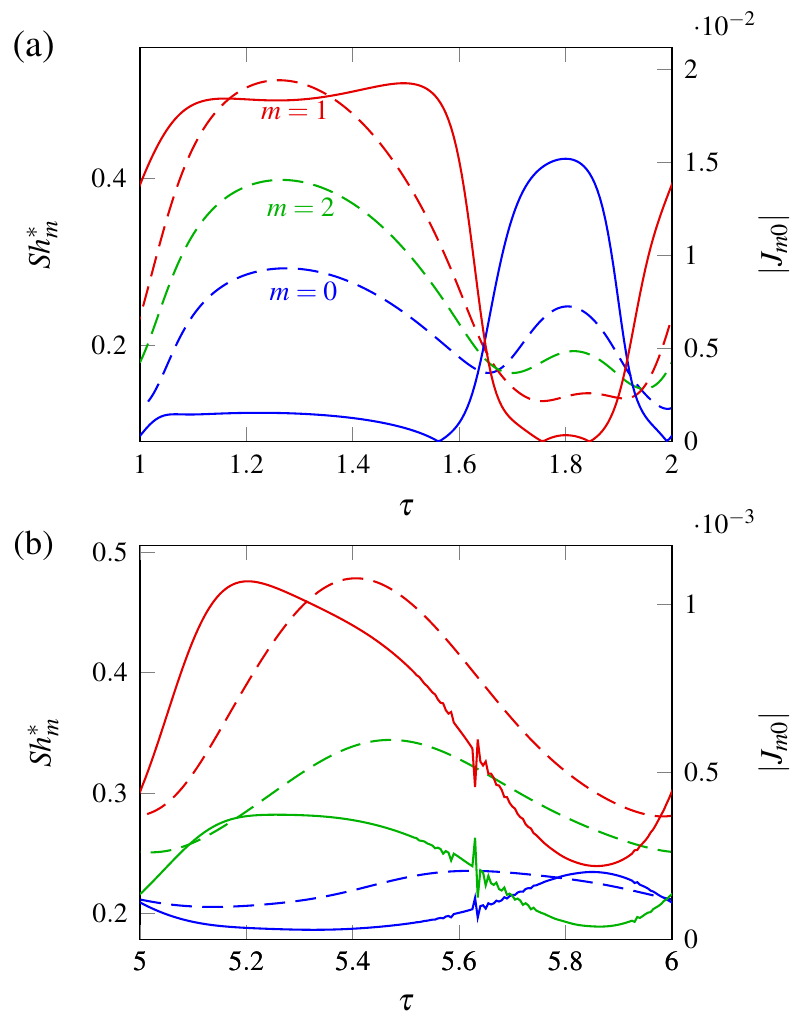}
	\caption[Sensitivites]{$\Sh^*_m$ (dashed lines) and absolute values of the $S$ sensitivity coefficients $J_{m0}$ (solid lines) for (a) case 1 and (b) case 2. Refer to Fig.~\ref{fig:5} for segments numbering. $J_{10}$ is not shown in (a) for visualization purposes.}
	\label{fig:6}
\end{figure}

\begin{figure*}
	\centering
	\includegraphics{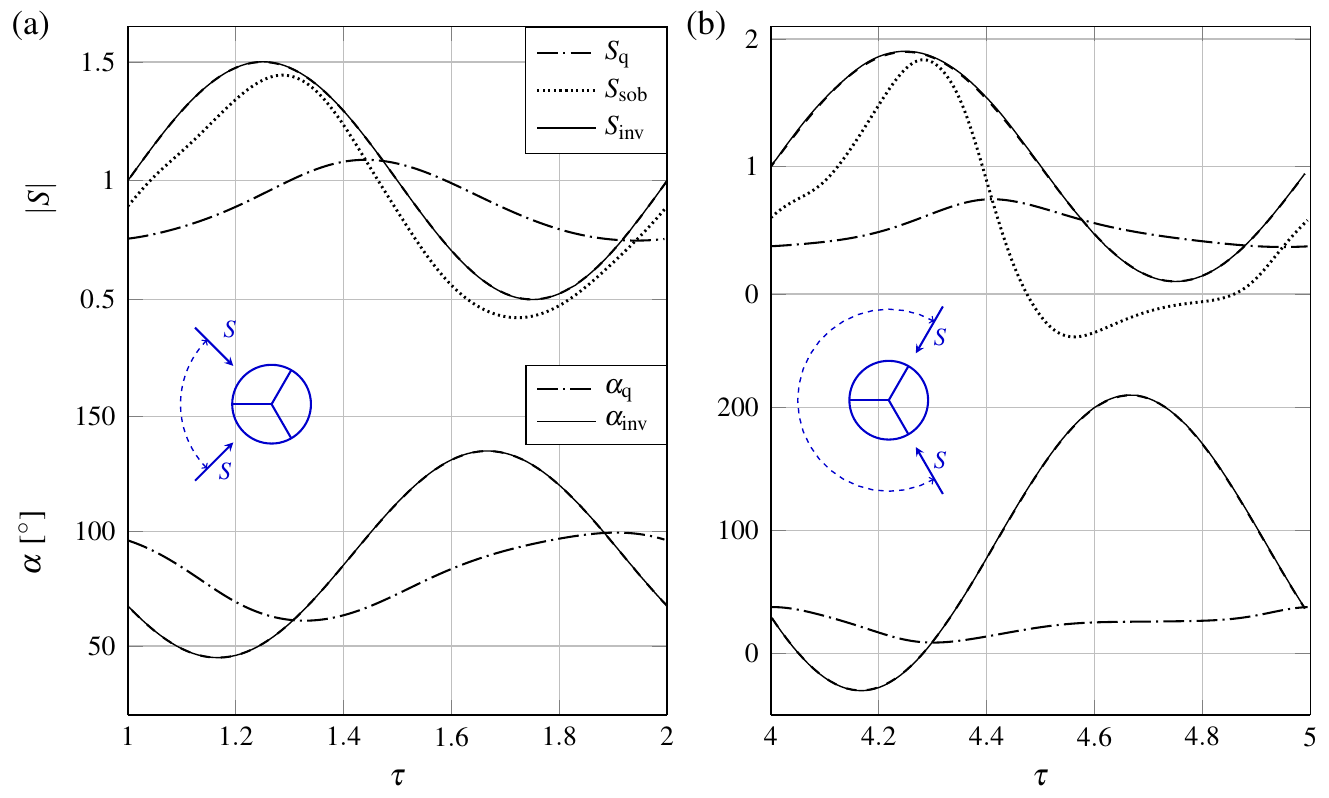}
	\caption[Inverse results 3]{Large amplitude, high-frequency fluctuations on both magnitude $S$ and direction $\alpha$ of the imposed wall shear rate, for (a) case 3 and (b) case 4 (see Table~\ref{tab:parnum}). Dashed lines indicate the true constraints.}
	\label{fig:7}
\end{figure*}

Evolutions of $\Sh^*_m(\tau)$ and $J_{m0}(\tau)$, the $S$ sensitivity coefficients, are plotted in Fig.~\ref{fig:6}. A rule of thumb is that the larger the $\Sh^*_m$, the higher this segment sensitivity will be, at least for quasi-steady processes. This is indeed observed for case 1 (Fig.~\ref{fig:6}a): as the shear reverses, the signal on the back segment $\Sh^*_0$ becomes the largest and so does its sensitivity magnitude $|J_{00}|$, which was essentially null for $\tau\lesssim1.6$. In the high-frequency case however (Fig.~\ref{fig:6}b), the duration of the shear reversal is too short considering the inertia of the process; the lagged sensitivity $|J_{00}|$ only slightly increases and $\Sh^*_0$ always stays with the lowest signal. Note that a similar trend is expected for the $\alpha$ sensitivity coefficients $J_{10}$ and $J_{12}$ (not shown here). Also notice the large drop of sensitivity between the two cases of Fig.~\ref{fig:6} (close to a factor $\sim10$), affecting the inverse process on both its stability and convergence speed as more iterations and time steps are needed for $\Sh^*_m$ to converge on $M_m$.  As a matter of fact, the more stable conjugated-gradient algorithm is required to procure the results shown for case 2; otherwise, oscillations like those observed on $\alpha$ after shear reversal (Fig.~\ref{fig:5}b) are more frequent and intense. The inverse method is also flawless when involving large amplitude fluctuations on both $S$ and $\alpha$ at high-frequency, as observed in Fig.~\ref{fig:7} for cases 3 and 4. Here again, quasi-steady and \sob methods exhibit strong departure from the imposed shear rate. Note that for the constant direction cases 1 and 2, the use of the three-segment probe is not essential as the problem becomes one-dimensional; the traditional sandwich probe (see Fig.~\ref{fig:1}b) is more convenient for treating shear reversal cases. Results shown although demonstrate that the proposed inverse method can deal with a very steep variation of the unknown variables such as the $\alpha(\tau)$ step-like signals in cases 1 and 2. 

\begin{figure}
	\centering
	\includegraphics{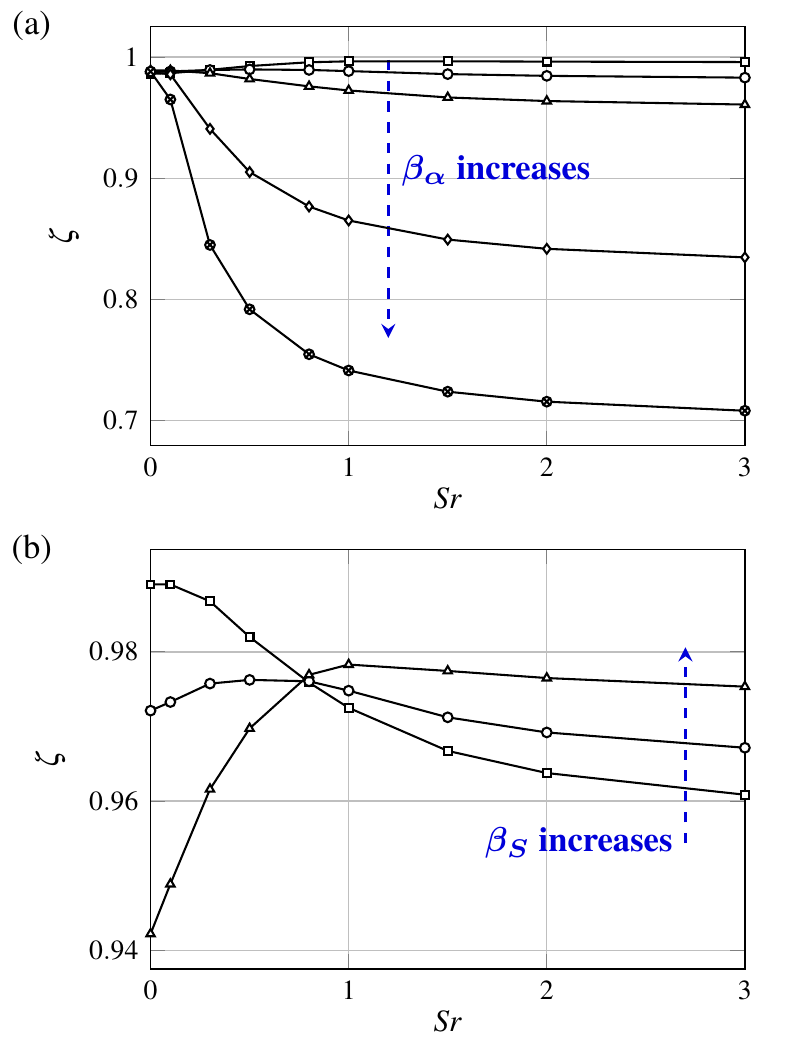}
	\caption[$\betaa$ effect on $\ol{\Sh}$]{Variation of $\zeta=\ol{\Sh^*}/\Sh^*_\text{std}$ with $\Sr$ for a constant amplitude (a) $\betas=0.5$ and (b) $\betaa=\pi/4$. Values are $\betaa=\{\pi/12,\,\pi/6,\,\pi/4,\,\pi/2,\,2\pi/3\}$ and $\betas=\{0.5,\,0.7,\,0.9\}$ in (a) and (b), respectively. $\zeta$ is calculated using the time averaged $\Sh^*$ over one period of the solicitation.}
	\label{fig:8}
\end{figure}

\begin{figure*}[h!]
	\centering
	\includegraphics{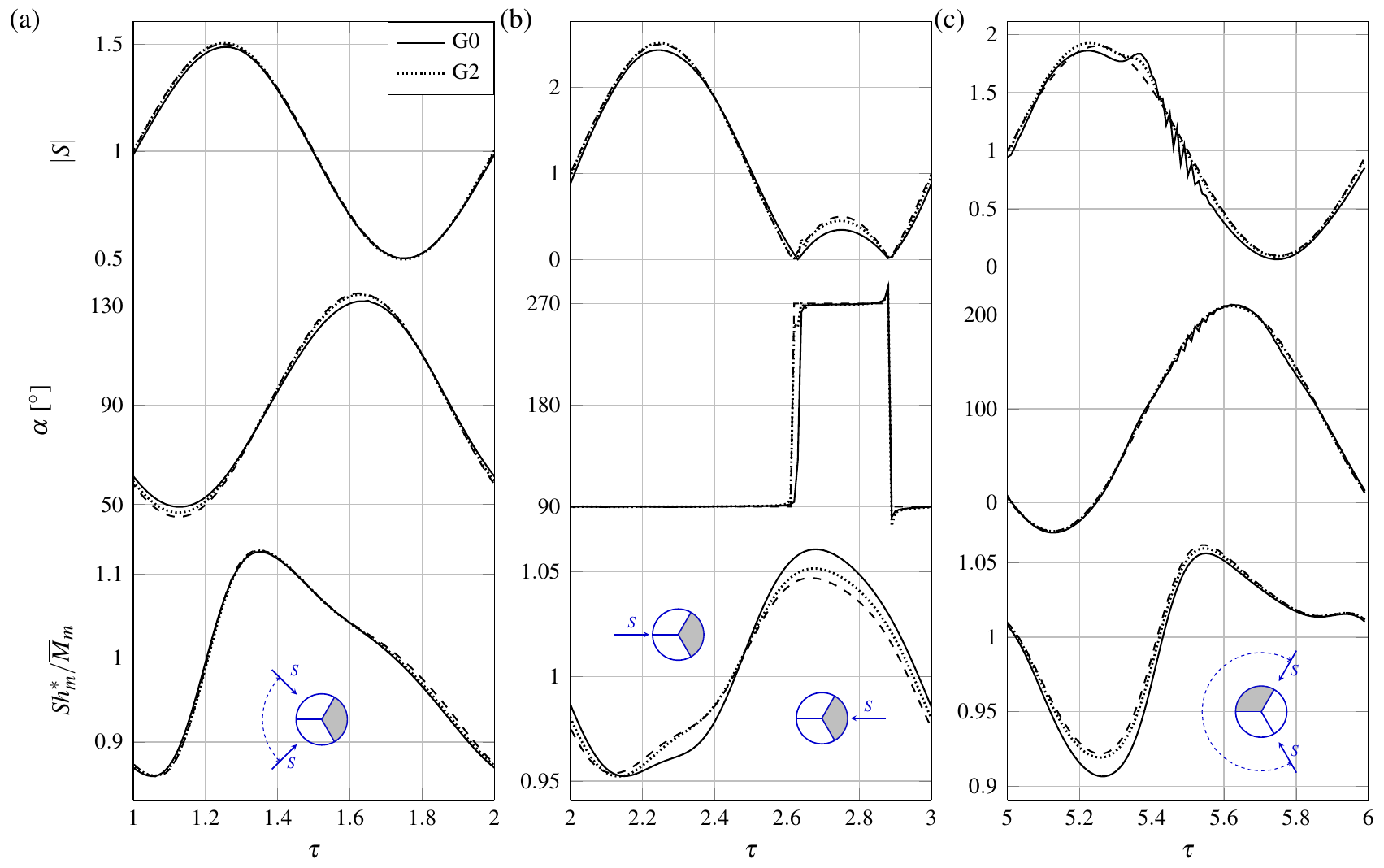}
	\caption[Gap effet on inverse results]{Effect of the probe geometry on the inverse method results for (a) case 5, (b) case 2 with $\alphao=\pi/2$ and (c) case 4, here with $\Pen=1\e{5}$. In all cases, results using G2 are better than with the null-gap geometry G0, yet offering a decent estimate of the true wall shear rate (illustrated with dashed lines). Phase lag and amplitude attenuation are nonetheless apparent, especially in the shear reversal period of case 2, justified in particular by the form of the filled segment $\Sh$ number (bottom figures). Using geometry G2 with large gaps, the convergence is considerably better. Note that the same effect occurs on the other segments, yet at a lesser extent. Results using G1 (not shown here) are very similar to those with G0.}
	\label{fig:10}
\end{figure*}

It is interesting to note that in both cases 3 and 4 the time average $\ol{\Sq}$ is shifted from the expected unitary value ($\ol{\Sq}\neq1$). Recalling $\eqref{eq:Shlev}$, $\Sq$ is calculated from the total Sherwood number $\Sh^*_\text{tot}=\sum\Sh^*_m$. As a result of the probe imperfections, the value of $k^*$ for mesh G1 (cf. Fig.~\ref{fig:9}a) differs from the theoretical null-gap geometry and is obtained by solving the stationary direct problem with $S=1$ and $\Pen\to\infty$, where in such case $k^*=\Sh^*_\text{tot}$. Although Fig.~\ref{fig:3}(b) exposes a slight variation of $\Sh^*_\text{tot}$ with $\alpha$ for a constant shear rate, this variation is not sufficient to explain the shift of $\ol{\Sq}$ as seen in Fig.~\ref{fig:7}. The offset actually arises considering that the ratio
\begin{equation}
	\zeta=\ol{\Sh^*}/\Sh^*_\text{std}<1
	\label{eq:zeta}
\end{equation}
for the unsteady case, with $\Sh^*_\text{std}$ the steady Sherwood number measured under corresponding conditions (i.e. same $\Pen$ and $\alphao$, $\Sr=0$). Fig.~\ref{fig:8} exposes this effect for a constant $\betas$ value in (a). At low $\Sr$, $\zeta$ is only slightly affected by variation of $\betaa$ whereas it largely depends on $\betas$ as observed in (b); values at $\Sr\to0$ actually tend to the one-dimensional case ($\betaa=0$) and are indeed fairly close to the ones obtained by \citet{kaip1983unst} for the one-dimensional shear rate case. For increasing values of $\betaa$ however, the $\zeta$ evolutions with $\Sr$ considerably differ since they do not converge to a value $\zeta=1$ at high $\Sr$\footnote{Tested up to $\Sr=50$ for the case $\betas=0.5$, $\betaa=\pi/4$, where $\zeta=0.957$.} as observed by \citeauthor{kaip1983unst}: the ratio decreases for larger $\betaa$ and smaller $\betas$. With $\zeta\neq1$, the steady calibration parameter $k^*$ then leads to an erroneous mean wall shear rate $\ol{\Sq}$, altering both $\Sq(\tau)$ and $\Ssob(\tau)$ as shown in Fig.~\ref{fig:7}. Note that while those results stand for the geometry of Fig.~\ref{fig:9}(a), the trends are similar for the null-gap geometry. $\zeta$ also varies only slightly with $\alphao$.

\subsection{Influence of the probe discretization \label{sec:probdis}}
When treating experimental data from a real three-segment sensor, one could use the actual geometry of the probe in the mesh construction in order to procure simulations as faithful as possible. Yet, taking into account the surface imperfections brings additional computational costs, both in the time committed to mesh generation and in the simulations themselves as more elements may be needed for a proper discretization. Respect of the gaps dimension in the simulated geometry might actually be the most important element in the process. To verify that statement, the two meshes of Figs.~\ref{fig:9}(b,c) were generated, namely G2 and G3. The geometry of the former consists in a perfect three-segment probe with regular gaps of dimension $g$ close to that of a real sensor; the latter is the real geometry itself, retrieved from a contour detection on an optical microscope photograph of the sensor (Fig.~\ref{fig:9}d). To simulate experimental conditions, mesh G3 was therefore used in the manner of Section~\ref{sec:valinv} to generate $M_m$ data using the direct problem for three different cases. Both meshes G0\footnote{As the inverse method results for meshes G0 (see Fig.~\ref{fig:2}c) and G1 are very similar, the latter are not presented. Also note that if one would use mesh G3 in the inverse problem, nearly perfect results are expected as the same mesh would be used in both direct and inverse problems.} and G2 were then used separately in the inverse problem to obtain the corrected time evolutions of $S$ and $\alpha$ corresponding to the simulated $M_m$. To account for the differences in $k^*$ exposed in Section~\ref{sec:valinv} for the different meshes, a scaling factor of the form $k^*_{\text{G}i}/k^*_\text{G3}$, $i\in\{0,2\}$, was applied on the  $M_m$ data. Using this procedure, the impact of the geometry is inspected. 

\begin{figure*}[th!]
	\centering
	\includegraphics[]{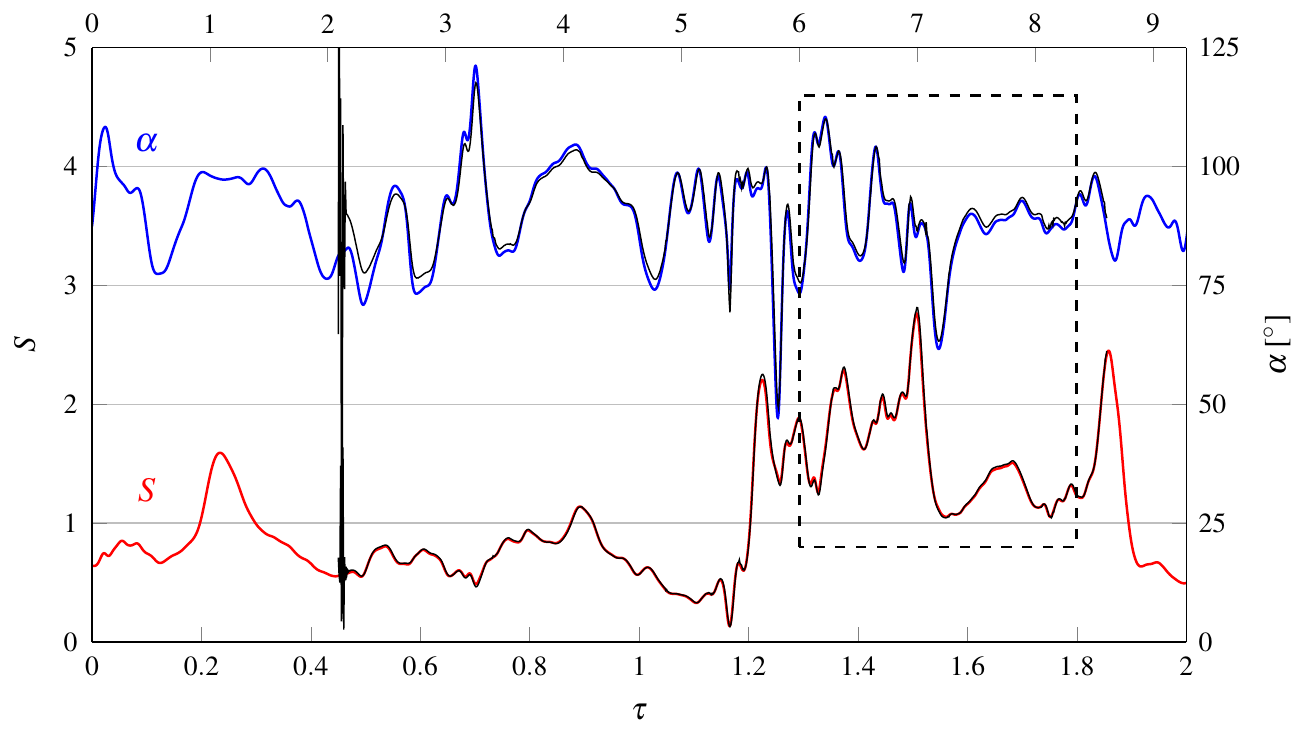}
	\caption{Two-component wall shear rate extracted from a DNS database of a turbulent Poiseuille flow at $\Rey=1\e{4}$ (colored curves). Results of the inverse method for the high $\Pen$ case are illustrated with thin black lines for both $S$ and $\alpha$. A close-up of the dashed rectangle region is presented in Figs.~\ref{fig:12} and \ref{fig:13}. Values for the dimensionless time $\tau=t^*\ol{s^*}\Pen^{-1/3}$ on the top and bottom abscissae concern the low and high $\Pen$ cases, respectively.}
	\label{fig:11}
\end{figure*}

Results for cases 5, 2 and 4 (here with $\alphao=0$ and $\Pen=1\e{5}$, see Table~\ref{tab:parnum}) are shown in Fig.~\ref{fig:10}. In all three examples, the mesh with realistic gaps G2 provides a better representation of $S$ and $\alpha$, in particular for case 2 near the shear reversal. As the proposed inverse method is based on minimizing the residuals $r_m$ (see Section~\ref{sec:invpro}) of all three segments, a compromise is unavoidable when the time evolution of the experimental data $M_m$ is complex like the one of case 2, characterized by two adjacent inflection points as seen in the lower Fig.~\ref{fig:10}(b). The absence of gaps in G0 cannot allow such evolution for the back segment Sherwood number and the best fit obtained is largely distorted. The convergence is somewhat better for the two other segments (not shown here), hence providing decent results for $S$ and $\alpha$. The major drawback is the phase lag and damping especially observable in the shear reversal period. With appropriate gaps, $\Sh^*_m(\tau)$ using G2 is remarkably more accurate, providing very good results for both $S$ and $\alpha$ even in the intense flow conditions of case 4 (Fig.~\ref{fig:10}c). The use of G0 here brings additional stabilization issues, manifested by sharp oscillations on $S$ and $\alpha$. Thus, for a real three-segment geometry without excessive imperfections like the one of Fig.~\ref{fig:9}(d), a perfect geometry with appropriate gaps would be an adequate trade-off between meshing complexity and inverse method results, although a mesh modeled on the actual geometry like G3 is always preferred when achievable. Still, Fig.~\ref{fig:10}(a) also illustrates that discrepancies using the null-gap geometry G0 could still be acceptable for many applications involving a two-dimensional oscillating shear rate.

\subsection{Non-periodic flow \label{sec:nonPer}}
\begin{figure*}[t]
	\centering
	\begin{minipage}{.48\textwidth}
		\centering
		\includegraphics[]{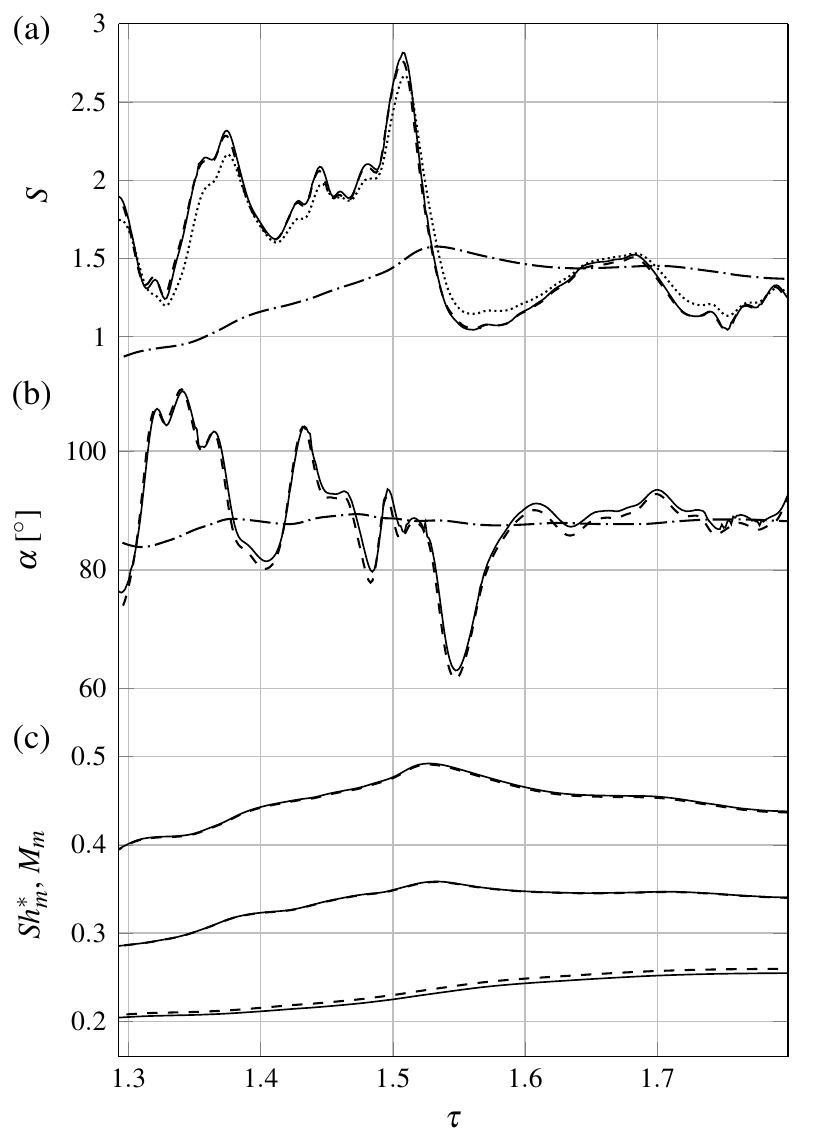}
		\caption{Results in the close-up region of Fig.~\ref{fig:11} for $\Pen=1.2\e{7}$. See Fig.~\ref{fig:7} for legend. Dashed lines indicate the true constraints.}
		\label{fig:12}
	\end{minipage}%
	\begin{minipage}{.04\textwidth}
			\hfill
	\end{minipage}%
	\begin{minipage}{.48\textwidth}
		\centering
		\includegraphics[]{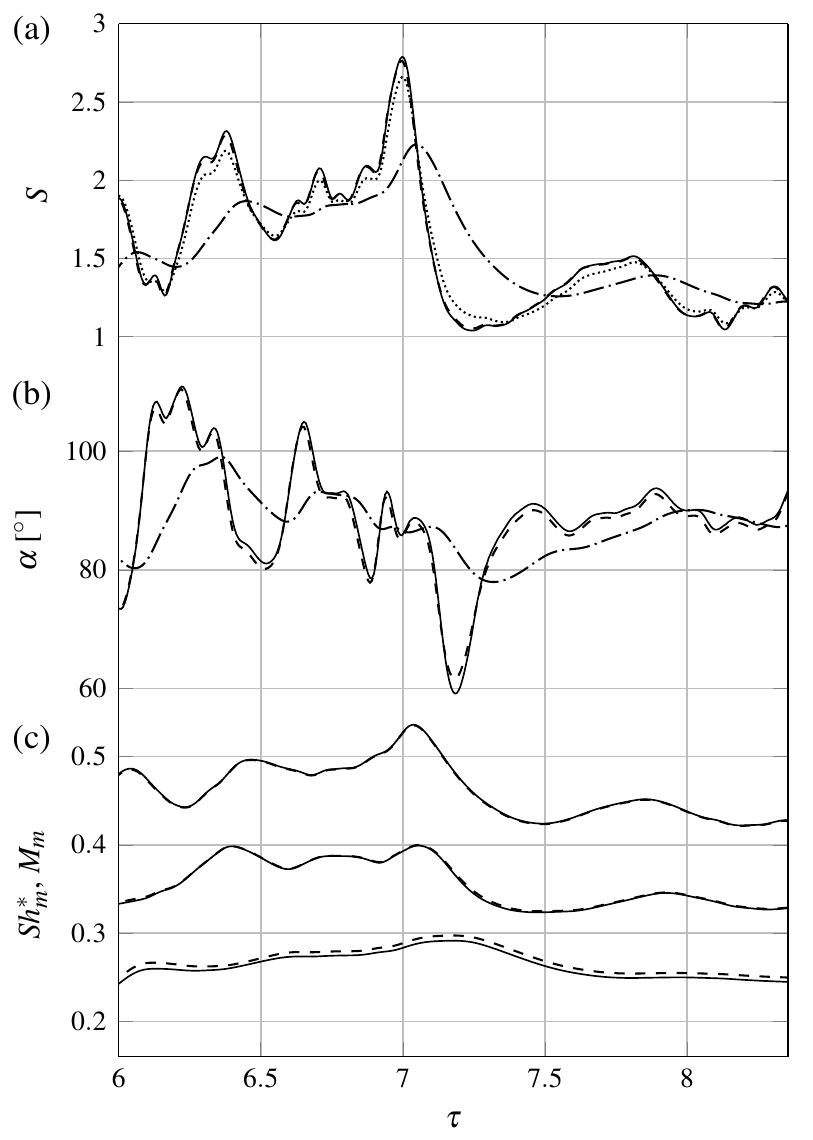}
		\caption{Results in the close-up region of Fig.~\ref{fig:11} for $\Pen=1.2\e{5}$. See Fig.~\ref{fig:7} for legend. Dashed lines indicate the true constraints.}
		\label{fig:13}
	\end{minipage}
\end{figure*}
While previous test cases were limited to periodic flows, the proposed inverse method can deal with stochastic variations of the wall shear rate magnitude and direction as those observed in turbulent flows. Hence, a direct numerical simulation (DNS) was performed\footnote{The finite difference solver \texttt{Incompact3d} \citep{incompact3d-0} was used for the DNS.} to generate a velocity database for the three-dimensional turbulent channel flow. The time evolution of the two-component wall shear rate was extracted at an arbitrary wall position (see Fig.~\ref{fig:11}) and then used to solve the direct problem and generate $M_m$ data for this turbulent flow.  

Instead of using a Strouhal number in the convection--diffusion and sensitivity equations, the dimensionless time in \eqref{eq:adim} is replaced with
\begin{equation}
	\tau = t\ol{s}\Pen^{-1/3}
	\label{eq:tauInst}
\end{equation}
or, using the dimensionless DNS variables (here represented with starred coefficients),
\begin{equation}
	\tau = t^*\ol{s^*}\Pen^{-1/3}.
	\label{eq:tauInst0}
\end{equation}
With such a parameter, a unitary coefficient then replaces the Strouhal number in front of the time derivative of equations \eqref{eq:cdadim3D} and \eqref{eq:C1C2} (equivalent to $\Sr=1$). $s^*$ was evaluated using finite difference approximation in the viscous sub-layer where a linear velocity profile was observed. Averaging was performed over the entire period shown in Fig.~\ref{fig:11} to evaluate $\ol{s^*}$. One can notice from \eqref{eq:tauInst0} that a higher $\Pen$ will decrease the resulting time step $\Delta\tau$ for the ED analysis, thus creating a more intense case for the inverse method considering its convergence properties are lessened with smaller time steps \citep{ozisik2000inverse}. Two values for the P\'eclet number were tested, namely $\Pen=1.2\e{7}$ and $\Pen=1.2\e{5}$, which respectively correspond to mean wall shear rates of \SI{36000}{\per\second} and \SI{360}{\per\second} provided that the typical values $d=0.5\,$mm and $D=7.5\e{-4}$\,$\si{\milli\meter\squared\per\second}$ are used \citep{sobolik1998calibration}. The Reynolds number $\Rey=Uh/\nu$ used in the DNS was $\Rey=1\e{4}$, with $U$, $h$ the average cross-sectional velocity and the channel height, respectively. 


To simulate experimental ED conditions, a procedure similar to that of Section~\ref{sec:probdis} was adopted. The $M_m$ data were first generated using mesh G3 while G2 was used to solve the inverse problem. One could see this additional complexity as the inevitable geometrical discordance between a real probe and the discretized one, considering for instance inactive areas on the sensor. Overall results for the high $\Pen$ case are presented in Fig.~\ref{fig:11} (where the inverse problem was started at $\tau=0.45$) while Figs.~\ref{fig:12} and \ref{fig:13} show a close-up view on the results for $\Pen=1.2\e{7}$ and $\Pen=1.2\e{5}$, respectively. As per  \eqref{eq:tauInst0}, the corresponding dimensionless time is $\Pen$ dependent; thus, the same DNS data is here characterized by two time evolutions (top and bottom abscissae in Fig.~\ref{fig:11}). A very good agreement is observed with the imposed wall shear rate for both $S$ and $\alpha$ in the two cases. The unstable character of the inverse method is well illustrated in Fig.~\ref{fig:11} by the large oscillations in the first tenth time steps, where boundness is ensured using a line search method; otherwise, the process would likely diverge. Time step size is also critical, where larger steps tend to stabilize the process and damp the oscillations. Only one out of four time steps from the simulated turbulent data was indeed used, which could explain the small discrepancies with the imposed solicitations. Note that the distinct meshes used in the data generation and inverse problem also inevitably introduces a certain degree of error, which most likely cannot be accounted for. This may also explain why in Fig.~\ref{fig:12}(c) an offset is observed between the lower $\Sh^*_m$ curve and its relative $M_m$. It is interesting to note the smooth and damped evolution of each segment signal $M_m$ in Fig.~\ref{fig:12}(c) for such unsteady flow, being a consequence of the severe inertia of the diffusion layer which, without usage of the inverse method, could hardly be accounted for. Such damping on the $M_m$ signals also justifies why the inverse method is so sensitive to noise; the use of filtering techniques thus appears to be essential when dealing with real experimental data. Besides, one can notice from Fig.~\ref{fig:12} that the straightforward \sob correction still procures acceptable results for $S(\tau)$, while the quasi-steady method is highly damped. At lower $\Pen$, one can notice from the $M_m$ curves in Fig.~\ref{fig:13}(c) that the probe is far more responsive to the imposed fluctuations; a slightly better agreement is also observed on the inverse results (Fig.~\ref{fig:13}a,b), although this effect is more apparent for the quasi-steady and, to a lesser extent, \sob methods. The reduced $\Pen$ also makes the inverse process less sensitive to noise and start-up instabilities.

\section{Concluding remarks}
An inverse problem algorithm coupled with the three-dimensional convection--diffusion equation is proposed in order to assess both magnitude and direction of the wall shear rate using electrodiffusion probes in high amplitude unsteady flows. The method was first validated and tested in flows of increasing complexity using simulated data. Results demonstrate that the inverse process not only surpasses all other post-treatment methods, but is the only valid one when dealing with shear reversal, periodically varying wall shear rate direction and turbulent flows, especially regarding the instantaneous shear direction. Numerical discretization of a real three-segment probe was also inspected. When the actual probe geometry is not accessible or easily discretized, a perfectly circular geometry with realistic interstices is suggested, which should offer satisfying results in most applications. Experimental work should be performed to complete the validation process and for further improvements of the two-dimensional inverse problem.

\section*{Acknowledgements}
The authors would like to acknowledge the financial support of the Natural Sciences and Engineering Research Council of Canada (NSERC) and the Fonds de recherche du Qu\'ebec - Nature et technologies (FRQNT). A most grateful thanks to Prof. V. Sobol\'ik, LaSIE Universit\'e de La Rochelle, for valuable advice and teaching of the electrodiffusion method and to Prof. A. Garon, Polytechnique Montreal, for the many recommendations regarding the finite element method.

\end{document}